\documentclass[%
reprint,
amsmath,amssymb,
aps,
]{revtex4-2}

\usepackage{graphicx}
\usepackage{dcolumn}
\usepackage{bm}

\begin{document}
	
	\preprint{APS/123-QED}
	
	\title{Evolution of the Fermi surface of 1T-VSe$_2$ across a structural phase transition}

\author{Turgut Yilmaz}
\affiliation{National Synchrotron Light Source II, Brookhaven National Lab, Upton, New York 11973, USA}
\email{tyilmaz@bnl.gov}

\author{Xiao Tong}
\affiliation{CenterforFunctionalNanomaterials,BrookhavenNationalLaboratory,Upton,NewYork11973,USA}

\author{Jerzy T. Sadowski}
\affiliation{CenterforFunctionalNanomaterials,BrookhavenNationalLaboratory,Upton,NewYork11973,USA}

\author{Sooyeon Hwang}
\affiliation{CenterforFunctionalNanomaterials,BrookhavenNationalLaboratory,Upton,NewYork11973,USA}

\author{Kenneth Evans-Lutterodt}
\affiliation{National Synchrotron Light Source II, Brookhaven National Lab, Upton, New York 11973, USA}

\author{Kim Kisslinger}
\affiliation{CenterforFunctionalNanomaterials,BrookhavenNationalLaboratory,Upton,NewYork11973,USA}

\author{Elio Vescovo}
\affiliation{National Synchrotron Light Source II, Brookhaven National Lab, Upton, New York 11973, USA}

\date{\today}

\begin{abstract}
The electronic origin of the structural transition in 1T-VSe$_2$ is re-evaluated through an extensive angle-resolved photoemission spectroscopy experiment. The components of the band structure, missing in previous reports, are revealed. Earlier observations, shown to be temperature independent and therefore not correlated with the phase transition, are explained in terms of the increased complexity of the band structure close to the Fermi level. Only the overall size of the Fermi surface is found to be positively correlated with the phase transition at 110 K. These observations, quite distant from the charge density wave scenario commonly considered for 1T-VSe$_2$, bring fresh perspectives toward the correct description of structural transitions in dichalcogenides materials.

\end{abstract}

\maketitle


\section{\label{sec:level1}Introduction\protect\\}

Two-dimensional (2D) layered transition metal dichalcogenides (TMDCs) display a variety of electronic, magnetic, and transport properties, making them extensively studied materials \cite{manzeli20172d, choi2017recent}. Among the many TMDCs, 1T-VSe$_2$ gained special attention as one of the few van der Waals (vdW) materials hosting a three-dimensional charge density wave (3D-CDW) phase \cite{eaglesham1986charge, tsutsumi1982x, coleman1988scanning, giambattista1990scanning}. The structural distortion sets in at a transition temperature (T$^{\ast}$) of 110 K, presenting a 4a $\times$ 4a $\times$ 3c periodic superlattice \cite{eaglesham1986charge, tsutsumi1982x, coleman1988scanning, giambattista1990scanning}. Electronically, the CDW phase manifests with a pseudogap at the Fermi level and a Fermi surface nesting along the CDW wave vector \cite{gruner1988dynamics}.

In spite of the apparently universal agreement on a CDW mechanism in VSe$_2$, its direct experimental evidence is far from compelling.
Early photoemission experiments from 1T-VSe$_2$ interpreted a shift of a secondary peak located just below the Fermi level as an indication of a gap opening \cite{terashima2003charge, sato2004three}, although a complete spectral weight suppression at the Fermi level was absent below T$^{\ast}$. In this approximation, the gap size was estimated between 20 meV to 50 meV. More recently, high-resolution angle-resolved photoemission spectroscopy (ARPES) experiments, however, did not report an energy gap around the M(L)-point of the Brillouin zone \cite{chen2022dimensional, chen2018unique, feng2018electronic}, where spectral distortions are expected to be stronger due to coincidence with the CDW wave vector. On the other hand, a small gap of 12 meV was detected by scanning tunneling spectroscopy (STM) \cite{jolie2019charge}. But, it was found to be more pronounced around the $\Gamma$-point rather than at $M$-point where the Fermi surface nesting is actually expected. Additional evidence for the CDW transition is a warping effect in the band dispersions along the k$_z$ direction of the 3D-Brillouin zone \cite{sato2004three, strocov2012three, wang2021three, coelho2019charge, kim2020dynamical, majchrzak2021switching}. Although this appears to be a well-established experimental result, it has never been tested for temperatures above T$^{\ast}$, making its relation to the CWD doubtful. In conclusion, the available experimental evidences present important inconsistencies with each other and with expected nature of a CDW phase.

Finally it may be worth mentioning that the presence of topological surfaces states has been recently revealed on the surface VSe$_2$\cite{Yilmaz2023}, implying a different Fermi surface from previous predictions and measurements. Furthermore resolving the Dirac cone in VSe$_2$ is not straightforward experiment, possibly explaining some of the inconsistencies in the previous studies. Thereby, the structural transition in VSe$_2$ has to be readdressed to investigate its impact on the electronic structure.

Here, we revisit the surface electronic structure of 1T-VSe$_2$ aiming to investigate the CDW phase. ARPES results virtually identical to the previous reports are reproduced using linear-horizontal (LH) polarized light. However, electronic states, not observed previously, are probed by linear-vertical (LV) polarized light. The latter bands contribute to the Fermi surface and need to be considered for an accurate description of the CDW phase. Their inclusion yields a distinct picture of the k$_z$ electronic dispersion, completing earlier band structure characterizations. In particular, the Fermi surface warping, usually attributed to the CDW phase, is found to be temperature-independent and results from the distinct k$_z$ dispersion of these multiple bands present at the Fermi level. On the other side, a prominent effect associated with the CDW phase is found in shrinkage of the electron pocket centered at $\overline{M'}$-point at 100 $\mp$ 5 K. This observation is in excellent agreement with transport measurements. Hence, the new findings in this work clarify many of the ambiguous observations present in the current literature while providing the band structure origin of the transport anomalies of 1T-VSe$_2$.

   \begin{figure*}[t]
	\centering
	\includegraphics[width=16cm,height=8.357cm]{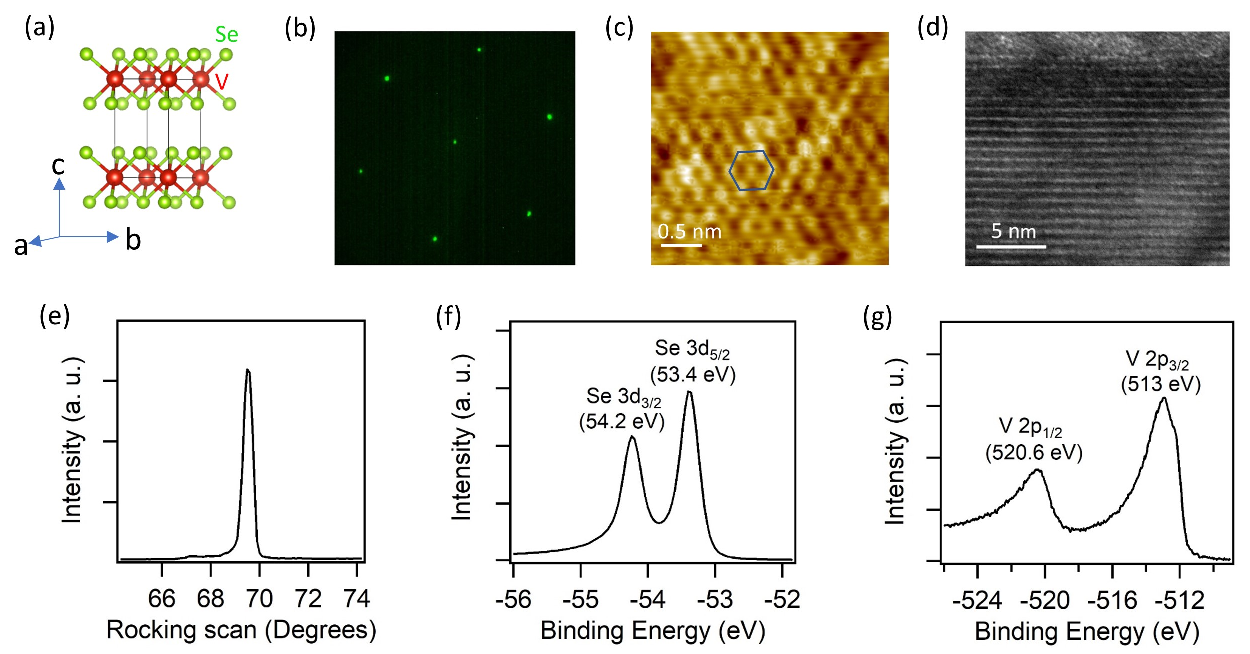}
	\caption{
		(a) Ball-stick representation of 1T-VSe2 crystal structure. (b-e) LEED pattern, STM image, HAADF-STEM cross-section image, and XRD pattern, respectively. (f-g) Se 3$d$ and V 2$p$ core levels spectra. All data were taken at room temperature except for the core levels recorded at 10 K. The blue hexagon in c represents the in plane unit cell. The STM image was recorded with a sample bias of 100 mV and set point of 1 nA. 
	}
\end{figure*}

   \begin{figure*}[t]
	\centering
	\includegraphics[width=16cm,height=8.543cm]{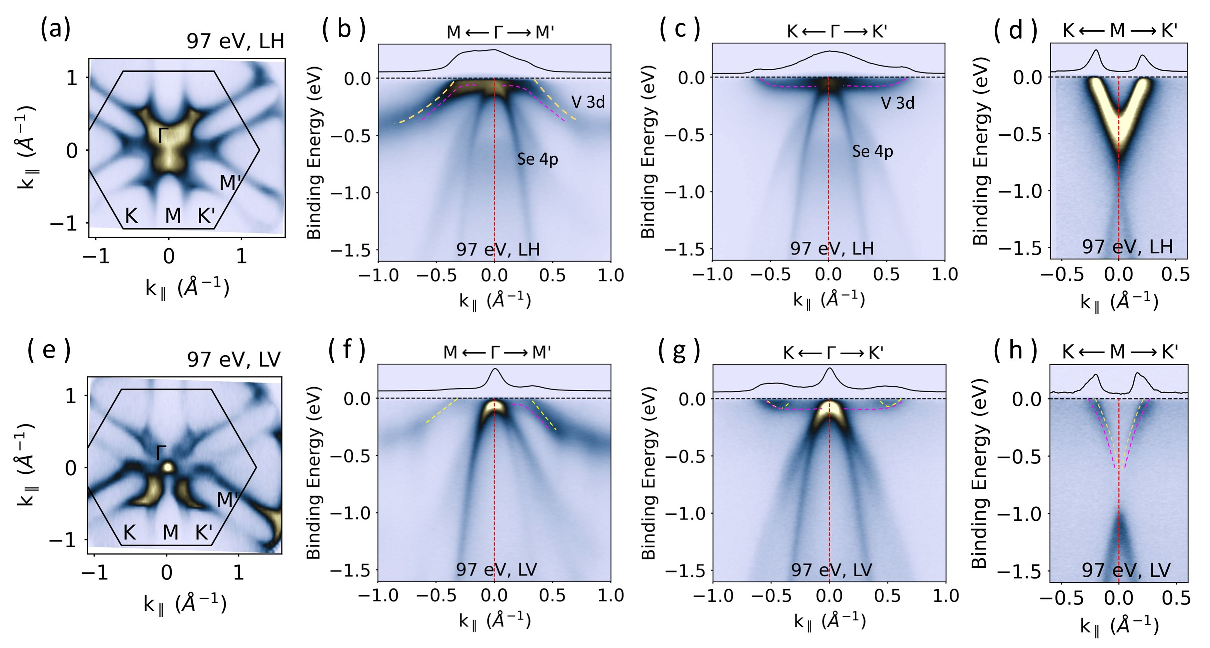}
	\caption{
		Top panels: LH polarization; bottom panels: LV polarization. (a) The Fermi surface. The superimposed hexagon represents the 2D-surface Brillouin zone. (b-d) ARPES spectra were taken along selected high symmetry directions indicated at the top of each spectrum. (e-h) same as (a-d) but for LV polarized light. The MDCs (black lines) integration is over 20 meV just below the Fermi level. All spectra were measured at 10 K with 97 eV photons corresponding to k$_z$ at the $\Gamma$-point. Pink and yellow dashed lines mark the bands close to the Fermi level.
	}
\end{figure*}

   \begin{figure*}[t]
	\centering
	\includegraphics[width=16cm,height=8.145cm]{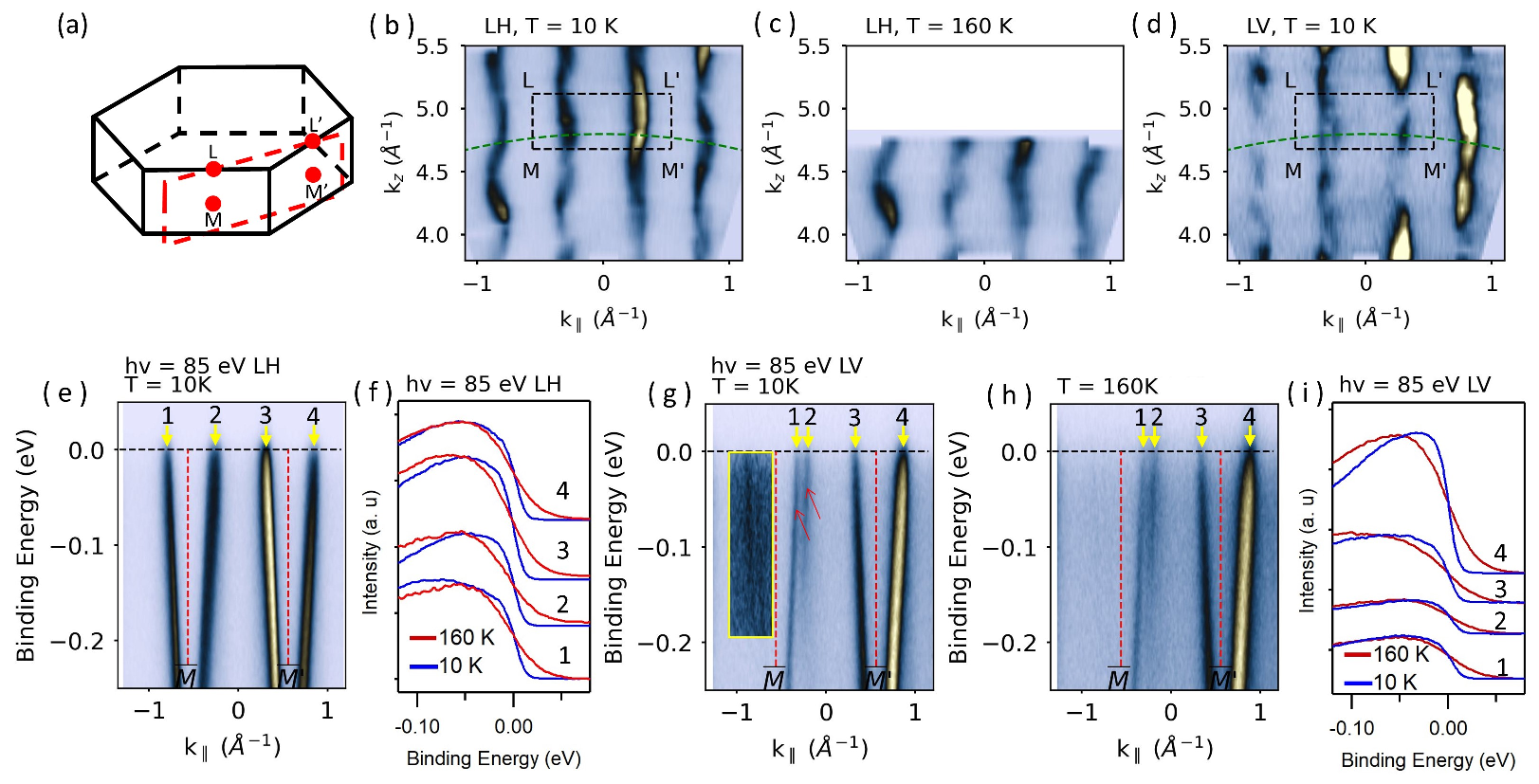}
	\caption{
		(a) Schematic representation of the 3D-Brillouin zone. (b-c) Fermi surface sections along the $\overline{MLL'M'}$ plane (see panel (a)) probed with LH light polarization at 10 K and 160 K, respectively. (d) Same as (b) but with LV polarized light. Green dashed arcs in (b) and (d) corresponds to 85 eV photon energy. (e) ARPES spectrum taken along the $\overline{M}$ - $\overline{M'}$ direction at 10 K with LH polarized light. (f) EDCs taken at 10 K (blue) and 160 K (red) at k$_{\parallel}$ marked with yellow arrows and number-labels in (e). (g-h) ARPES spectra along the $\overline{M}$ - $\overline{M'}$ direction taken with LV light polarization at 10 K and 160 K, respectively. (i) EDCs taken at 10 K (blue) and 160 K (red) at k$_{\parallel}$ marked with yellow arrows in (g-h). Red arrows in (g) mark the multiple bands. The yellow rectangle highlights an area where the color has been saturated to make the bands visible.
	}
\end{figure*}

   \begin{figure*}[t]
	\centering
	\includegraphics[width=16cm,height=9.821cm]{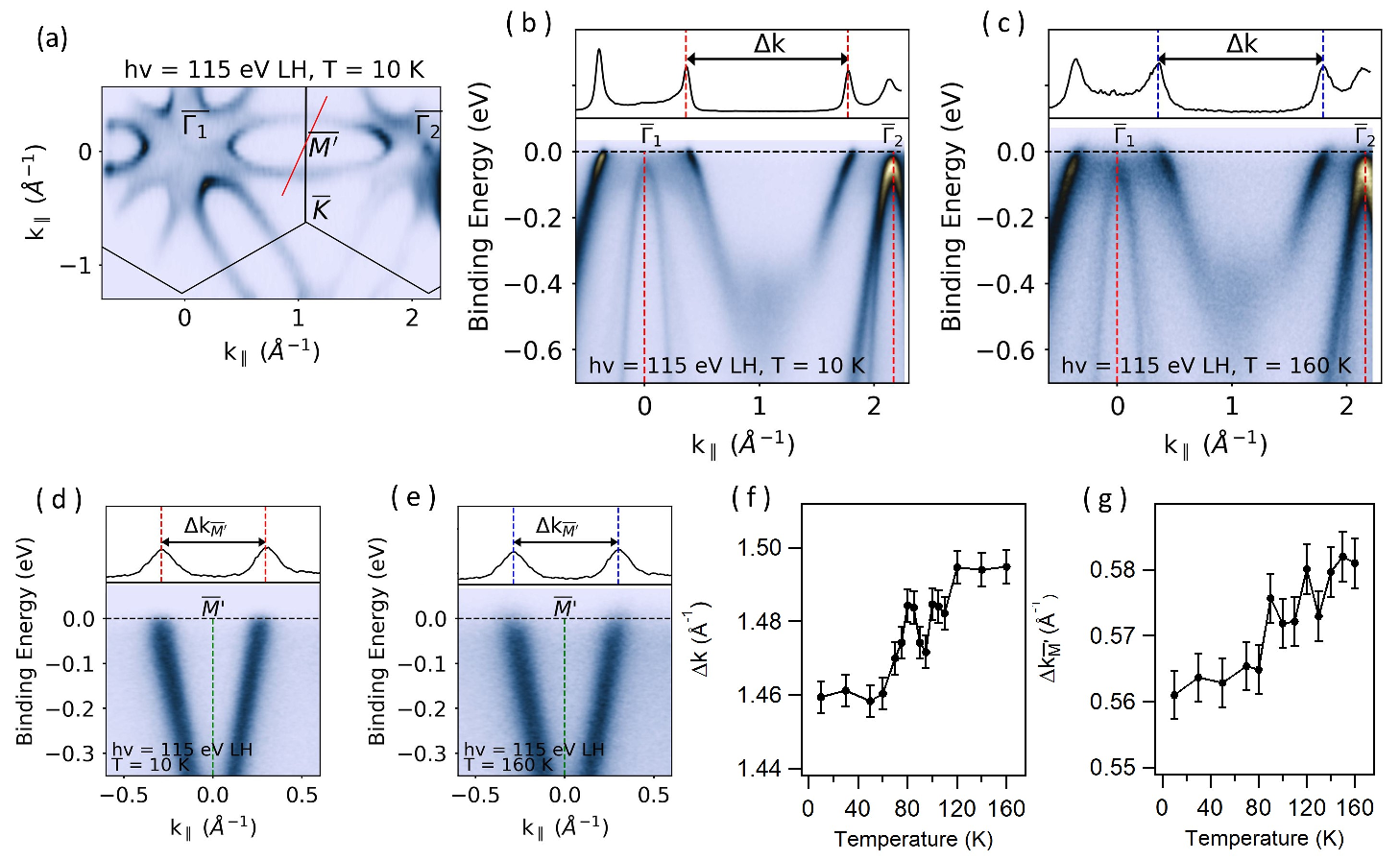}
	\caption{
		(a) Fermi surface taken at 10 K with 115 eV LH polarized light. (b-c) ARPES spectra along the $\overline{\Gamma}_1$ - $\overline{\Gamma}_2$ direction at 10 K and 160 K, respectively. (d-e) ARPES spectra taken along the red line in a at 10 K and 160 K, respectively. MDCs taken at the Fermi level are given at the top of each spectrum. (f-g) $\Delta$k and $\Delta$k$_{\overline{M'}}$ as a function of sample temperature, respectively. These quantities are determined by fitting the peaks in the MDCs with Lorentzian functions on linear backgrounds.
	}
\end{figure*}

\section{\label{sec:level2}EXPERIMENTAL METHODS\protect\\}

Single crystal 1T-VSe$_2$ samples are prepared through floating zone method. ARPES and core level experiments were performed at 21-ID-1 (ESM) beamline of NSLS-II using a DA30 Scienta electron spectrometer. The base pressure in the photoemission chamber was 1x10$^{-11}$ Torr. The energy resolution was better than 12 meV and the beam spot size was approximately 5 $\mu$$^2$. The synchrotron radiation incidence angle was 55$^o$. Analyzer slit was along to the $M$ - $\Gamma$ - $M'$ direction during the ARPES measurements at normal emission and parallel to the $M$ - $M'$ direction during the zone corner scan. Linear verticle (LV) polarized light is parallel to the sample surface and analyzer slit while linear horizontal (LH) polarized light is on the incident plane..

High-angle annular dark-field imaging scanning tunneling electron microscopy (HAADF-STEM) images were acquired with Hitachi HD2700C dedicated STEM with a probe Cs corrector operating at 200 kV at room temperature. TEM lamella were prepared using the in-situ lift-out method on the Helios 600 NanoLab DualBeam FIB, with final Ga$+$ milling performed at 2 keV. Scanning tunneling microscopy (STM) (omicron VT-STM-XA 650) experiments were performed in an ultrahigh vacuum system with a base pressure of 2x10$^{-10}$ Torr at room temperature. The STM images were observed in constant current mode using Pt/Ir tips. TEM and STM images were analyzed by using Gwyddion-2.55 software package. HAADF-STEM and STM experiments were conducted at the Center for Functional Nanomaterials, Brookhaven National Laboratory.
Micro low energy electron difraction ($\mu$LEED) experiment was performed at X-ray photoemission electron microscopy/low-energy electron microscopy (XPEEM/LEEM) endstation of the ESM beamline (21-ID-2).
X-ray diffraction (XRD) experiment was carried out on the ISR beamline, 4-ID, with a wavelength of 1.079 $\AA$. A Dectris 1M area detector was used with a Huber six circle diffractometer to obtain detailed reciprocal space maps of the crystal.

\section{\label{sec:level3}EXPERIMENTAL RESULTS\protect\\}

The crystal structure of 1T-VSe$_2$ is examined by multiple techniques to confirm the high quality of the samples and consistency with earlier studies. A ball-stick representation of the atomic structure is shown in Figure 1(a). 1T-VSe$_2$ crystallizes with an in-plane hexagonal geometry as shown in micro-low energy electron diffraction (µLEED) and STM images given in Figure 1(b-c). Both data indicate the presence of a single domain phase without extra diffraction spots or other in-plane atomic arrangements. A representative high-resolution tunneling electron microscope (TEM) image further confirms the layered structure along the c-axis of the crystal with well-defined atomic layers and smooth interfaces between the mono-layers (Figure 1(d)). X-ray diffraction (XRD) data in Figure 1(e) assigns in-plane and out-plane lattice constants of 3.354 $\AA$ and 6.097 $\AA$ respectively, in excellent agreement with the literature5. The rocking curve of the high-quality single crystal was 0.4 degrees full width at half maximum with negligible secondary crystal grains. Finally, the chemical environment of the sample is probed by Se 3$d$ and V 2$p$ core level spectra which show no additional spectral features beyond the spin-orbit splitting components (Figure 1(f-g)). The binding energies correspond to -2 and +4 oxidation states for Se and V atoms, consistent with a single chemical phase in the material \cite{liu2018epitaxially}.

The surface electronic structure of a 1T-VSe$_2$ bulk sample is shown in Figure 2 for LH (upper panel) and LV (lower panel) polarized lights. The LH Fermi surface exhibits ellipsoidal pockets centered at the {M}($M'$)-points and an intense spectral feature in the zone center due to Se 4$p$-atomic orbitals (Figure 2(a)). The electronic structure along the $M$ - $\Gamma$ - $M'$ direction looks identical to the previous reports \cite{terashima2003charge, sato2004three, chen2022dimensional, chen2018unique, feng2018electronic, jolie2019charge, strocov2012three, wang2021three, coelho2019charge, kim2020dynamical, majchrzak2021switching}. But the Fermi level energy distribution curve (EDC) given above the spectrum shows weak shoulders indicating the presence of two distinct bands close to the Fermi level (see dashed pink and yellow lines in Figure 2(b)). One, resembling the band structure calculations \cite{kim2020dynamical, majchrzak2021switching}, bends into a flat dispersion towards the $\Gamma$-point and overlaps with the Se 4$p$ derived bands. The other continues instead the upward dispersion, crossing the Fermi level before reaching the center of the zone.

The electronic structure along the $K$ - $\Gamma$ - $K'$ direction is shown in Figure 2(c). The V-3d derived bands start with a flat dispersion in the vicinity of the zone center and bend upward, crossing the Fermi level at k$_{\parallel}$ = -0.6 $\AA$$^{-1}$ and 0.65 $\AA$$^{-1}$. Furthermore, an intense V-shaped band characterizes the region along the $K$ - $\Gamma$ - $K'$ direction (Figure 1(d)). This is the main region of interest for the CDW, where the gap is expected to be more prominent.

The spectra taken with LV polarization are presented in the bottom panels of Figure 2. The main difference between the Fermi surfaces with LH and LV polarization is that now only a point-like feature is observed at the $\Gamma$-point, facilitating the analysis of the energy-momentum maps. Along the $M$ - $\Gamma$ - $M'$ direction, the crossing of the bands at the Fermi level is now more easily identified. Furthermore, along the $K$ - $\Gamma$ - $K'$ direction, two small electron pockets (marked with yellow dashed lines) are additionally resolved at k$_{\parallel}$ = -0.45 $\AA$$^{-1}$ and k$_{\parallel}$ = 0.54 $\AA$$^{-1}$, respectively. More importantly, the ARPES spectrum taken along the $K$ - $\Gamma$ - $K'$ direction shows multiple bands, marked with dashed red and yellow lines, crossing the Fermi level (Figure 2(h)). They can be better distinguished in the momentum distribution curve (MDC) given at the top of the spectrum. The observation of the multiple bands touching the Fermi level around the M-point is a novelty of the present study. Therefore, this work provides a crucial component of the electronic structure to draw a correct picture of the CDW phase in 1T-VSe$_2$.

Up to now, the most solid evidence for the 3D-CDW phase in 1T-VSe$_2$ is given by the warping effect on the Fermi surface along the k$_z$ direction \cite{sato2004three, strocov2012three, wang2021three}. However, the observation of distinct states with LH and LV polarized lights suggests to re-investigate this issue with both polarizations. It is also worth emphasizing that previous photon energy-dependent ARPES experiments were not performed as a function of temperature, clearly a must-do step to ascertain the electronic origin of the phase transition. Thereby, a such experiment is presented in Figure 3. Figure 3(b-d) shows the k$_z$ dispersion for states at the Fermi level in the $\overline{MLL'M'}$) plane measured with LH polarization. At 10 K, far below the T$^{\ast}$, the results are very similar to the one in the literature. In particular, the pronounced asymmetries between the right and left sides around the $M(L)$-points - the so-called warping effect - were attributed to the CDW-induced Fermi surface nesting \cite{sato2004three, strocov2012three, wang2021three}.  However, the k$_z$-dispersion taken at 160 K, well above the T$^{\ast}$, is essentially identical to the 10 K one, featuring the same 'warping' along the k$_z$-direction (Figure 3(c)). These modulations of the spectral intensity or dispersion, therefore, cannot be attributed to the phase transition and their origin must be searched elsewhere. A hint comes from the dispersion map constructed with LV polarized light (Figure 3(d)). This reveals a more complex picture, featuring additional bands, best captured in the splitting resolved at certain photon energies. The observed asymmetric dispersion behavior in the photon energy dependencies can therefore be simply explained by the presence of multiple states close to the Fermi level combined with their photon energy-dependent photoemission cross-sections.

The above findings also suggest that the CDW-induced gap should be studied with both LH and LV polarization. For LH polarization, the ARPES spectrum taken at 85 eV along the $\overline{M}$ - $\overline{M'}$ direction is given in Figure 3(j) together with selected EDCs taken below (10 K) and above (160 K) the T$^{\ast}$ in Figure 3(f).  Clearly, the EDCs exhibit no gap opening at the Fermi level in this case. Similar experiments for LV polarization are shown in Figure 3(g-i). Although multiple bands are captured close to the Fermi level with LV polarization, neither of them displays a prominent gap at the Fermi level (Figure 3(i)).

The detailed band structure characterizations given above convincingly demonstrate that the previous observations stem from multi-band crossing the Fermi level rather than the CDW phase transition. This implies that the impact of the structural distortion on the electronic structure of 1T-VSe$_2$ remains an open question. On the other end, although the Fermi surface does not display gaps and nesting based on the ARPES experiments, transport measurements show clear anomalies around 110 K \cite{van1976magnetic, thompson1979correlated, li2020structural, ortenzi2009fermi}. In some cases, structural transitions can cause modifications of the Fermi surface area \cite{coldea2019evolution, fernandes2014drives, watson2016evidence, fuh2016newtype}. Therefore, it seems sensible to investigate the size of the ellipsoidal electron pockets centered at $\overline{M'}$-point as a function of temperature. A Fermi surface portion covering two Brillouin zone centers is given in Figure 4(a). Here, two relevant quantities are $\Delta$k, measuring the distance between the ellipsoidal pockets along the $\Gamma$$_1$ - $\Gamma$$_2$ direction, and $\Delta$k$_{\overline{M'}}$, measuring the size of the pocket along the red line shown in Figure 4(a). Representative spectra below and above T$^{\ast}$ are given in Figure 4(b-c) and (d-e), where $\Delta$k and $\Delta$k$_{\overline{M'}}$ are also marked. Plots of these two quantities as a function of temperature are presented in Figure 4(f-g). Interestingly, both, $\Delta$k and $\Delta$k$_{\overline{M'}}$ undergo a sizable change across the CDW temperature, indicating a decrease in the Fermi surface area with the temperature decreasing from 160 K or 10 K. This result is in excellent agreement with the transport measurements which show a relative increase of the in-plane resistivity across the same temperature \cite{van1976magnetic, thompson1979correlated, li2020structural}. Thereby, this finding establishes a direct connection between the electronic structure and the structural distortion in 1T-VSe$_2$.

\section{\label{sec:level4}CONCLUSION and DISCUSSION\protect\\}

A detailed study of the electronic structure of 1T-VSe$_2$ has been conducted to assess the electronic origin of the CDW phase. The new ARPES data reveals the presence of previously undetected states located in the vicinity of the Fermi level and therefore, necessarily relevant to the physical properties of 1T-VSe$_2$. However, no gap opening at the Fermi level is observed in correspondence with the CDW temperature. Furthermore, previous claims of a connection between the Fermi surface profile along the k$_z$ direction and a 3D-CDW in 1T-VSe$_2$ are also ruled out. On the other hand, a change in the size of the in-plane Fermi surface across the CDW transition is firmly established, in good correspondence with the transport results. 

A remaining issue in our work is to determine to what extent this Fermi surface 'breathing' is or should be put in relation to the structural transition. Either a shift of the bands or a change in the effective masses would cause such Fermi surface modifications. Based on our initial studies and analyses, both are playing role. Unfortunately, the broadening of the bands at higher binding energies and the weak modifications observed in the electronic structure, prevent to arrive at an accurate conclusion at this point. However, one striking perspective of the current observation is their possible connection with analogous Fermi surface shrink commonly observed in Fe-based superconductors \cite{ortenzi2009fermi, coldea2019evolution, fernandes2014drives, watson2016evidence}. In those cases, it is believed that a nematic transition pushes the bands up or down depending on their orbital character, leading to a band splitting and consequent Fermi surface modifications \cite{fuh2016newtype}. In this case, it may be possible to drive 1T-VSe$_2$ into a superconducting state by playing with its Fermi surface shrinking. Indeed, recent observations on the superconducting properties of the 1T-VSe$_2$ under high pressure \cite{sahoo2020pressure} or growth conditions \cite{yilmaz2022spectroscopic} could be related to the results presented here.

\begin{acknowledgments}
This research used resources ESM (21ID-I) beamline of the National Synchrotron Light Source II, a U.S. Department of Energy (DOE) Office of Science User Facility operated for the DOE Office of Science by Brookhaven National Laboratory under Contract No. DE-SC0012704. We have no conflict of interest, financial or other to declare.
\end{acknowledgments}

\end{document}